\documentstyle[12pt]{article}
\newcommand{\bl}{\Bigl(}
\newcommand{\br}{\Bigr)}
\newcommand{\cd}{c^{\dagger}}
\newcommand{\be}{\beta}
\newcommand{\eps}{\epsilon}
\newcommand{\g}{\gamma}
\newcommand{\s}{\sigma}
\newcommand{\la}{\lambda}
\newcommand{\La}{\Lambda}
\newcommand{\th}{\theta}
\newcommand{\lan}{\lambda_{\alpha}^{n}}
\newcommand{\lgm}{\lambda_{\gamma}^{m}}

\newcommand{\half}{{1\over 2}}
\newcommand{\beq}{\begin{equation}}
\newcommand{\eeq}{\end{equation}}
\newcommand{\up}{\uparrow}
\newcommand{\down}{\downarrow}

\begin{document}
\begin{flushright}
ITP-SB-92-72
\end{flushright}

\begin{center}

\vskip3em
{\bf \huge Thermodynamics and Excitations of the Supersymmetric
$t-J$ Model}

\vskip3em
{\large Eric D. Williams\footnote{e-mail: williams@max.physics.sunysb.edu}}

\vskip1em

{Institute for Theoretical Physics, SUNY, Stony Brook 11794}

\end{center}

\vskip 2cm

\begin{abstract}
The free energy of the supersymmetric $t-J$ model is
expressed in terms of finite temperature excitations above
thermodynamic equilibrium.  This reveals that the free energy has the
form of noninteracting fermions with temperature dependant excitation
spectra. We also discuss the ground state and zero temperature
excitations.
\end{abstract}

\section{Introduction}
\setcounter{equation}{0}

Strong electron correlations are believed to be important in
understanding the high $T_{c}$ superconductors~\cite{anderson1}.  This
idea is supported by the fact that the high $T_{c}$ compounds display
antiferromagnetism in the absence of doping.  The $t-J$ model
describes strongly correlated electrons with antiferromagnetic
exchange interactions, and has been proposed as a candidate to model
high $T_{c}$ superconductors by Zhang and Rice~\cite{zhang}.  The one
dimensional version of the model becomes integrable~\cite{schlottman}
and supersymmetric~\cite{super} for special values of the parameters
$t$ and $J$.  Integrability allows the model to be solved by Bethe's
ansatz and supersymmetry leads to the construction of three distinct
Bethe ansatz solutions.  Lai~\cite{lai} and
Sutherland~\cite{sutherland} each found a solution in the context of
models of hard core bosons and fermions which are now understood to be
equivalent to the supersymmetric $t-J$ model.
Schlottman~\cite{schlottman} applied the Bethe ansatz to the $t-J$
hamiltonian, and found Lai's solution describes the model when
electrons are treated as dynamical objects in a background of empty
sites.  Sarkar~\cite{graded} discovered that Sutherland's solution
applies to the $t-J$ model if one treats holes and spin down electrons
as dynamical objects in a background of spin up electrons.  There is a
third and quite recent solution due to Essler and
Korepin~\cite{essler}, which is similar to Sutherland's, but
interchanges the roles of the spin down electrons and the holes.

The ground state of the $t-J$ model was constructed
in~\cite{schlottman,lai,graded}.  The free energy was calculated in
{}~\cite{schlottman}. Bares, Blatter, and Ogata~\cite{bares} use both
Lai's and Sutherland's ansatzes to give a detailed account of the
ground state and excitation spectrum, including extensive numerical
analysis.

The outline of the paper is as follows: in sec.~2 we define the model
and take the thermodynamic limit of the the Bethe ansatz equations.
In sec.~3, the bulk free energy is expressed in terms of
finite temperature excitations.  In these terms, the free energy
has the form of a noninteracting system with temperature dependant
excitation spectra.  This structure has been observed in such models
as the $\delta$-function bose gas~\cite{yang} and the XXZ spin
chain~\cite{johnson}, here these ideas are extended to th	e $t-J$ model.
In sec.~4 we discuss the ground state and the order one excitation
spectrum. The results agree with those of~\cite{schlottman,bares}, we
present them here for completeness, and to see how see how they arise
from a different form of Bethe ansatz.  In the appendix we note that the
different ansatzes are related by an interchange of particle and hole
rapidities.

\section{Formulation}
 \setcounter{equation}{0}

The $t-J$ model describes spin-$\half$ fermions on a lattice.  The
on-site coulomb repulsion is taken to be infinite, so that there is no
double occupancy of the sites.  The constraint of no double occupancy
implies that the number of empty sites (holes) is conserved, which
allows one to think of the empty sites as dynamical objects. The
parameter $t$ describes how easily the electrons hop from site to site
and $J$ describes the strength of an antiferromagnetic exchange
interaction.  The Hamiltonian is
\beq
\label{hamil}
    H={\cal P} \{ \sum_{j=1}^{L} \sum_{\s=\up,\down} t(\cd_{j,\s} c_{j+1,\s}
 +\cd_{j+1,\s} c_{j,\s}) + J (\vec{S}_{j} \cdot \vec{S}_{j+1}
-{1 \over 4} n_{j} n_{j+1}) \} {\cal P} +2 \hat{N} -L.
\eeq
$L$ is the number of lattice sites. $ c_{j,\s}$, and
$c_{j',\s'}^{\dagger}$ are electron creation and annihilation operators
which obey the anti-commutation relations
\beq
 \{c_{j,\s},c_{j',\s'} \}=0 , \ \ \
  \{ c_{j,\s},c_{j',\s'}^{\dagger} \}= \delta_{j,j'}\delta_{\s,\s'} ,
     \ \ \s,\s'= \up,\down.
\eeq
$\cal{P}$ is the operator which projects out doubly occupied sites,
$\vec{S}_{j} = \cd_{j,\s}
\vec{\sigma}_{\s,\s'} c_{j,\s'}$, where $
\vec{\sigma}_{\s,\s'}$ is the vector of Pauli spin matrices, and
$n_{j}$ is the number of electrons at site $j$. $\hat{N}$ is the
operator for the total number of electrons.  We have added the term $2
\hat{N} -L$ onto the definition of the Hamiltonian
in~\cite{schlottman,bares}, which just shifts the energy and chemical
potential.  We shall consider this model at the special point
$J=2t=2$, where the model is integrable and
supersymmetric~\cite{super}.  Also, at this special point the
Hamiltonian~(\ref{hamil}) may be rewritten as a graded permutation
operator~\cite{graded},
\beq
          H = - \sum_{j=1}^{L} \Pi_{j,j+1},
\eeq
where $\Pi_{j,j+1}$ interchanges the states on neighboring
sites, with a minus sign if electrons sit on both sites.

The eigenstates and spectrum of the Hamiltonian~(\ref{hamil}) at the
supersymmetric point may be found by nested Bethe ansatz, a technique
introduced by Yang in~\cite{yang2}.  The spectrum of the
Hamiltonian~(\ref{hamil}) and other conserved quantities are given in
terms of a set of coupled algebraic equations known as Bethe ansatz
equations(BAE).  As mentioned in the introduction, there are three
distinct sets of BAE for the model.  We shall analyze the solution due
to Essler and Korepin~\cite{essler}, where the coordinates of holes
and spin down electrons are taken as dynamical objects moving in a
background of spin up electrons. This solution expresses the
eigenvalues of the Hamiltonian and total momentum as:
\beq
             E = L - \sum_{l=1}^{N_{\down}+N_{h}}
                 \frac{4}{\la_{l}^{2} + 1},
\eeq
\beq
              P= \sum_{l=1}^{N_{\down}+N_{h}}
              i \log \bl{\frac{\la_{l}+i}{\la_{l}-i}} \br,
\eeq
where the $N_{h}+N_{\down}$ spectral parameters $\la_{l}$ describe the
motion of holes and spin down electrons.  They must satisfy the Bethe
ansatz equations
\beq
     \label{pbc1}
     {\bl \frac{\la_{l} +i}{\la_{l} -i} \br}^{L} =
          \prod_{\beta =1}^{N_{\down}}
    \frac{\la_{l} - \La_{\beta} +i}{\la_{l}-\La_{\beta}-i}
\ \ \ \ \ l=1,...,N_{\down}+N_{h},
\eeq
\beq
\label{pbc2}
        1= \prod_{l=1}^{N_{\down} + N_{h}}
      \frac{\la_{l} - \La_{\beta} +i}{\la_{l}-\La_{\beta}-i}
\ \ \ \ \ \beta=1,...,N_{\down}.
\eeq
$N_{\down}$, $N_{\up}$, and $N_{h} = L - N_{\down}-N_{\up}$  are
the number of spin up electrons, spin down electrons, and holes
respectively.  The $N_{\down}$ parameters $\La_{\beta}$ describe the
motion of the spin down electrons relative to the holes.

The equations~(\ref{pbc1}) and~(\ref{pbc2}) have real and complex
solutions.  The complex solutions are of a special form known as
strings, which may be found by fixing $N_{h}$ and $N_{\down}$ and
letting the lattice size $L$ go to infinity. One finds complex
solutions consisting of $n$ $\la$'s and $n-1$ $\La$'s in the following
combinations~\cite{essler}:
\begin{eqnarray}
\label{string}
       \la_{j} & = & \la + i(n+1 - 2 j) \qquad j=1,...,n  \nonumber \\
     \La_{\delta} & = & \la + i(n-2 \delta) \qquad \ \  \delta = 1,...,n-1,
\end{eqnarray}
for arbitrary $n=1,...,\infty$. The real part common to all the
parameters, $\la$, is known as the center of the string.  Physically,
string solutions are bound states in the sense that an eigenfunction
with the string combinations decays exponentially with respect to the
coordinates of the down spins and holes.  We shall see later in this
section that the string solutions ~(\ref{string}) yield a complete
spectrum of the hamiltonian. In a finite box, the string
solutions~(\ref{string}) are not exact.  We assume, along with many
authors (ie.~\cite{takahashi,gaudin,lai2}), that the corrections vanish in the
thermodynamic limit.

We wish to write BAE for the centers of the strings.  Let
$\la_{\alpha}^{n}$ the center of the $\alpha$th string of length $n$,
and $\La_{\beta}$ be real rapidities for spin down electrons not in
any bound state.  Denote the number of strings of length $n$ by
$N_{n}$ and the number of free spin downs by $N_{\La}$. Multiplying
together the equations for each part of the string yields
\beq
   \label{pbcs1}
   \bl \frac{\la_{\alpha}^{n} + in}{\la_{\alpha}^{n} - in}\br ^{L} =
     \prod_{m=1}^{\infty}
 \prod_{\stackrel{{\scriptstyle \g=1}}
 {{\scriptscriptstyle (m,\g) \neq (n,\alpha)}}}^{N_{n}}
F_{nm}(\la_{\alpha}^{n} - \la_{\g}^{m})
          \prod_{\be=1}^{N_{\La}}
 \frac{\la_{\alpha}^{n}-\La_{\be} + i n}{\la_{\alpha}^{n}-\La_{\be} - i n}
\eeq
where $n=1,...,\infty$ and
\beq
\label{pbcs2}
1= \prod_{n=1}^{\infty} \prod_{\alpha=1}^{N_{n}}
\frac{\La_{\beta} -\la_{\alpha}^{n} +i n}{\La_{\beta} -\la_{\alpha}^{n} - in},
\eeq
where
\beq
     F_{nm}(x) = e(\frac{x}{|n-m|})e^{2}(\frac{x}{|n-m|+2})
         \times \cdots \times e^{2}(\frac{x}{n+m-2}),
\eeq
and we have defined $e(x) =\frac{x+i}{x-i}$.

The energy and momentum in terms of the centers of the strings are
\beq
         E= L - \sum_{n=1}^{\infty} \sum_{\alpha=1}^{N_{n}}
           \frac{4n}{(\lan)^{2} + n^{2}},
\eeq
and
\beq
         P= \sum_{n=1}^{\infty} \sum_{\alpha=1}^{N_{n}}
                  p(\frac{\lan}{n}),
\eeq
where $p(\la)= 2\tan^{-1}(\la) - \pi$.

In order to count the solutions and take the thermodynamic
limit of the BAE, we take the logarithm of~(\ref{pbcs1}) and
{}~(\ref{pbcs2}):
\beq
     L\theta(\frac{\lan}{n}) =
      2 \pi I_{\alpha}^{n} + \sum_{m=1}^{\infty} \sum_{\g=1}^{N_{m}}
      \Theta_{nm}(\lan - \la_{\g}^{m}) + \sum_{\be = 1}^{N_{\La}}
      \theta(\frac{\lan -\La_{\be}}{n}),
\eeq
and
\beq
          \sum_{n=1}^{\infty} \sum_{\alpha}^{N_{n}}
          \theta(\frac{\La_{\be} - \lan}{n})
           = 2 \pi J_{\be},
\eeq
where $\theta(x) = 2 \tan^{-1}(x)$,
and
\beq
\Theta_{nm}(x) = \cases{\th(\frac{x}{|n-m|})
+ 2\th(\frac{x}{|n-m|+2})
         + \cdots + 2\th(\frac{x}{n+m-2})  & if $n \neq m $ \cr
       2\th(\frac{x}{2})+2\th(\frac{x}{4})+
        \cdots +2\th(\frac{x}{n+m-2}) & if $n=m$. \cr}
\eeq
 $I_{\alpha}^{n}$ and $J_{\beta}$ are integers (half integers) arising
from the choice of the branch of the logarithm.  $I_{\alpha}^{n}$ is
an integer (half integer) if $(L-\sum_{m \neq n}
N_{m} - N_{\La})$ is even (odd).  $J_{\beta}$ is an integer (half
integer) if $(\sum_{n=1}^{\infty} N_{n})$ is even
(odd).

Given a solution $\{ \lan \},\{ \La_{\beta} \}$ define
functions
\beq
\label{z1}
         z^{n}(\la) \equiv  \theta(\frac{\la}{n}) -
         \frac{1}{L} \sum_{m=1}^{\infty} \sum_{\g=1}^{N_{m}}
           \Theta_{nm} ( \la - \lgm) -
     \frac{1}{L}  \sum_{\be=1}^{N_{\La}} \theta(\frac{\la-\La_{\be}}{n})
\eeq
for $n=1,...,\infty$, and
\beq
\label{z2}
   z^{\La}(\La) = \frac{1}{L}
 \sum_{n=1}^{\infty} \sum_{\alpha=1}^{N_{n}} \theta(\frac{\La-\lan}{n}).
\eeq
By definition, $L z^{n}(\lan) = 2 \pi I^{n}_{\alpha}$ and $L
z^{\La}(\La_{\beta}) = 2 \pi J_{\beta}$.  If $z^{n}(\la)$ and
$z^{\La}(\La)$ are monotonic functions, then  specifying
sets of integers $\{I_{\alpha}^{n}\}$ and $\{J_{\beta}\}$ uniquely
determines the solution of (\ref{pbcs1}) and (\ref{pbcs2}).  Define
the integers corresponding to infinite spectral parameter by $2 \pi
I_{\infty}^{n} = L z^{n}(\infty)$ and $2\pi J_{\infty} = L
z^{\La}(\infty)$. In order to count the states correctly, we discard
solutions with infinite spectral parameter, yielding the maximum
allowable integers
\begin{eqnarray}
\label{max1}
    I_{max}^{n} &= & I_{\infty}^{n} -1 \nonumber \\
                &=& \half ( L - \sum_{m=1}^{\infty} N_{m}(t_{nm}-1)
                  -N_{\La} -2),
\end{eqnarray}
and
\begin{eqnarray}
\label{max2}
    J_{max} &=& J_{\infty} -1  \\ \nonumber
            &=& \half (\sum_{n=1}^{\infty} N_{n} - 2),
\end{eqnarray}
where $t_{nm} = 2 {\rm min}(m,n) - \delta_{nm}$.  Counting solutions
of the BAE (\ref{pbcs1}) and (\ref{pbcs2}) using the above, and taking
into account that each Bethe state generates a supersymmetry multiplet
of size $d=8 S_{z}$, one obtains a sum for the total number of states
identical to the corresponding formula found by Foerster and Karowski
using Sutherland's ansatz in~\cite{foerster}. There it is shown that
the sum adds up to $3^{L}$, the total number of possible states. Thus
there are no additional complex solutions other than~(\ref{string})
contributing to the spectrum.

We may also use equations ~(\ref{z1}) and ~(\ref{z2}) to define hole
rapidities.  Having assumed that $z^{n}$ and $z^{\La}$ are monotonic
functions of $\la$ and $\La$, then for any {\it unoccupied} integer
$\bar{I}^{n}_{\alpha}$, there exist a corresponding hole rapidity
$\bar{\la}_{\alpha}^{n}$.  We will call the combined sets of
$\{\lan\}$ and $\{\bar{\la}_{\alpha}^{n}\}$ the {\it vacancies} for
parameters.  Vacancy integers run through the entire allowed range
of integers.

\subsection{Thermodynamic Limit of BAE}

We now take the thermodynamic limit of the Bethe ansatz
equations~(\ref{pbcs1}) and~(\ref{pbcs2}), fixing the electron and
magnetization densities.   The solutions to the BAE become densely
packed with differences between neighboring $\la_{j+1}^{n}-\la_{j}^{n},
\La_{j+1}- \La_{j} \sim {\cal O}(1/L)$.  One passes to a description of
the states in terms of densities.  Define particle and hole densities:
\beq
        \rho_{n}^{p} (\la)\equiv \lim_{L \rightarrow \infty}
          \frac{1}{L(\la_{I_{j+1}^{n}} - \la_{I_{j}^{n}})},\ \ \ \
        \rho_{n}^{h} (\la) \equiv \lim_{L \rightarrow \infty}
          \frac{1}{L(\bar{\la}_{\bar{I}_{j+1}^{n}}
          - \bar{\la}_{\bar{I}_{j}^{n}})}
\eeq
and
\beq
         \sigma^{p} (\La) \equiv \lim_{L \rightarrow \infty}
         \frac{1}{L(\La_{J_{j+1}} - \La_{J_{j}})},\ \ \ \
         \sigma^{h} (\La) \equiv \lim_{L \rightarrow \infty}
         \frac{1}{L(\bar{\La}_{\bar{J}_{j+1}} - \bar{\La}_{\bar{J}_{j}})},
\eeq
and the total density of vacancies is $\rho_{n}^{t} = \rho_{n}^{p}
+\rho_{n}^{h}$ and $\sigma_{t} = \s^{p} + \s^{h}$. In the
thermodynamic limit, the Bethe ansatz equations become integral
relations between the densities:
\beq
\label{tpbc1}
       \rho_{n}^{t}(\la) = f_{n}(\la) - \sum_{m=1}^{\infty}
  \int_{-\infty}^{\infty} d\la' A_{nm}(\la-\la') \rho^{p}_{m}(\la')
    - \int_{-\infty}^{\infty} d\La f_{n} (\la - \La) \s^{p}(\La)
\eeq
and
\beq
\label{tpbc2}
       \s^{t} (\La) = \sum_{n=1}^{\infty} \int_{-\infty}^{\infty}
  d\la f_{n} (\La - \la) \rho_{n}^{p} (\la),
\eeq
where
\beq
      f_{n} (\la) = {1 \over 2 \pi} \frac{d}{d\la} \theta(\la)
                  = {1 \over \pi} \frac{n}{\la^{2} + n^{2}}
\eeq
and
\beq
     A_{nm} (\la) =  \cases{
      f_{|n-m|}(\la) + 2 f_{|n-m|+2} + \cdots + 2f_{n+m-2} & if $n \neq m$ \cr
           2 f_{2} + 2 f_{4} + \cdots + 2f_{2n-2} & if $n=m$. \cr }
\eeq

\section{Thermodynamics}
\subsection{Free Energy}
\setcounter{equation}{0}

To calculate the free energy, we follow the techniques developed by
Yang and Yang~\cite{yang}, Takahashi~\cite{takahashi}, and
Gaudin~\cite{gaudin}.  The free energy is a functional of the
densities:
\beq
          F = E  -\mu N - B S_{z} -TS,
\eeq
where the energy is (all integrals beyond this point are between
$-\infty$ and $\infty$ unless explicitly otherwise)
\beq
      E= L(1 - \sum_{n=1}^{\infty} \int d\la
             4 \pi f_{n} (\la) \rho_{n}^{p} (\la)),
\eeq
the number of particles is
\beq
\label{density}
      N = L(1- \int d\la \sum_{n=1}^{\infty} \rho_{n}^{p}(\la) +
            \int d\La \s^{p}(\La)),
\eeq
the magnetization is
\beq
       2 S_{z} = L(1- \sum_{n=1}^{\infty} (2n-1) \int d\la \rho_{n}^{p}(\la)
           - \int d\La \s^{p}(\La)),
\eeq
and the entropy is given by
\beq
 \label{entropy}
   S= L \Bigl[\int d\la  \sum_{n=1}^{\infty} [\rho_{n}^{t} \log(\rho_{n}^{t})
          - \rho_{n}^{p} \log (\rho_{n}^{p}) -\rho_{n}^{h} \log(\rho_{n}^{h})]
  +\int d\La[\s^{t} \log(\s^{t}) - \s^{p} \log(\s^{p}) -\s^{h} \log(\s^{h})]
  \Bigr].
\eeq
To obtain the entropy~(\ref{entropy}), note that the number of states in
in an interval $d\la$ in one of the fermi seas is
\beq
        N(\la,d\la) = e^{S(\la) d\la} =
          \frac{[L \rho^{t} d\la]!}{[L\rho^{p} d\la]! [L \rho^{h} d\la]!}.
\eeq
This the number of ways to put $[L \rho^{p} d\la]$ ``particles'' into
$[L \rho^{t} d\la]$ vacancies.  Using Stirling's approximations for
the factorial function at large $N$,$\log N! \sim N \log N $, and
adding up the contribution due to all types of particles, one arrives at
{}~(\ref{entropy}).

We now have the free energy for an arbitrary state.  To find the
equilibrium value, one minimizes the free energy with respect to the
densities.  We take the $\rho_{n}^{p}$ and $\s^{p}$ as the independent
variables, with $\rho^{t}_{n}$ and $\s^{t}$ determined from
{}~(\ref{tpbc1}) and ~(\ref{tpbc2}). Setting the variation of the free
energy equal to zero yields a infinite set of coupled nonlinear integral
equations for the equilibrium densities:
\beq
\label{te1}
    \eps_{n}(\la) = \eps^{0}_{n}(\la) + T \sum_{m=1}^{\infty} \int d\nu
     A_{nm} (\la-\nu) \log (1+ e^{- {\eps_{m}(\nu) \over T}})
     -T \int d\La f_{n}(\la-\La) \log(1+ e^{-{\eps_{\La}(\La) \over T}}),
\eeq
with $n=1,...,\infty$, and
\beq
\label{te2}
     \eps_{\La}(\La) = -\mu -B + T \sum_{n=1}^{\infty} \int d\la
       f_{n} (\La - \la) \log(1+e^{-{\eps_{n}(\la) \over T}}),
\eeq
where we have written the equations in terms of the functions
\beq
     \eps_{n} \equiv T \log({\rho_{n}^{h} \over \rho_{n}^{p}}),\ \ \ \
     \eps_{\La} \equiv T \log ({\s^{h} \over \s^{p}}),
\eeq
and
\beq
\label{energy2}
      \eps^{0}_{n}(\la) = -4 \pi f_{n}(\la) - B(2n-1) +\mu.
\eeq
Evaluating the free energy at the minimum yields the equilibrium value
\begin{eqnarray}
\label{free}
      \frac{F}{L}  & = & (1+B-\mu)
-T \sum_{n=1}^{\infty}\int d\la
      f_{n}(\la) \log(1+e^{-{\eps_{n}(\la) \over T}}).
  \\ & = & -1 -T \int d\la f_{1}(\la) \log(1+e^{-{\eps_{1}(\la) \over T}})
       - T \int d\La f_{2}(\La) \log(1+e^{-{\eps_{\La}(\La) \over T}}) \\
          & = &  (1-2 \mu) - \eps_{\La}(\La=0).
\end{eqnarray}
The density as a function of the chemical potential is determined from~
(\ref{density}).  The expressions (\ref{te1}),({\ref{te2}), and (\ref{free})
determining the free energy are equivalent to the
corresponding expressions in~\cite{schlottman} as indicated in
the appendix.

\subsection{Finite Temperature Excitations}


We now construct finite temperature excitations above the equilibrium
state, following the techniques developed in~\cite{yang}.  The idea is
as follows: The equilibrium state consists of an infinite number of
fermi seas.  The state is highly degenerate, so arbitrarily choose one
representative.  Now add a particle to one of the seas, all the other
seas will change in response.  The energy of such an excitation is
independent of which equilibrium state one started from, and is called
a finite temperature excitation.

 We begin with a single particle excitation in the sea of strings of
length $n'$.  Consider an equilibrium state for large but finite $L$.
If one inserts an additional particle into the sea at $\la_{p}$, this
leads to an order one excitation above the equilibrium state.  This is
shown in Figure 1. The rapidities in all the seas will
shift, $\lan \rightarrow \tilde{\la}_{\alpha}^{n}$, $\La_{\be}
\rightarrow \tilde{\La}_{\be} $, with $\la_\alpha^{n}
-\tilde{\la}_{\alpha}^{n} \sim \La_{\be}-
\tilde{\La}_{\be} \sim  {\cal O} (1/L)$.
 We will describe this flow by introducing shift functions~\cite{korepin}:
\beq
      S_{n}(\la|\la_{p}) \equiv \lim_{L \rightarrow \infty}
          \frac{\lan - \tilde{\la}_{\alpha}^{n}}{\la_{\alpha +1}^{n} - \lan},
\eeq
 and
\beq
      S_{\La}(\La|\la_{p}) \equiv \lim_{L \rightarrow \infty}
  \frac{\La_{\be} - \tilde{\La}_{\beta}}{\La_{\be +1} - \La_{\be}},
\eeq
where the difference in the denominator is between neighboring
{\it vacancies}. The energy of the excitation is
\begin{eqnarray}
      \Delta E & = & \eps^{0}_{n'}(\la_{p})
       + \sum_{n=1}^{\infty} \sum_{\alpha=1}^{N_{n}}
  (\eps^{0}_{n}(\tilde{\la}_{\alpha}^{n}) - \eps^{0}_{n}(\lan)) \\
               & \simeq & \eps^{0}_{n'}(\la_{p})
        - \sum_{n=1}^{\infty} \sum_{\alpha=1}^{N_{n}}
(\partial_{\la} \eps^{0}_{n})(\lan)
           (\lan - \tilde{\la}_{\alpha}^{n}).
\end{eqnarray}
In the thermodynamic limit the sums become integrals
\beq
       \Delta E =  \eps^{0}_{n'}(\la_{p})
      - \sum_{n=1}^{\infty} \int d\la  (\partial_{\la}\eps^{0}_{n})(\la)
       \theta_{n}(\la) S_{n}(\la|\la_{p}),
\eeq
where $\theta_{n}(\la) = (1+e^{\frac{\eps_{n}}{T}})^{-1}$ and
$\th_{\La} = (1+e^{\frac{\eps_{\La}}{T}})^{-1}$. In a similar way,
the momentum of the excitation is determined to be
\beq
\label{pn}
       \Delta P = p(\frac{\la_{p}}{n'})
  -\sum_{n=1}^{\infty} \int p'(\frac{\la}{n}) \th_{n}(\la)
      S_{n}(\la|\la_{p}) \equiv k_{n}(\la_{p}),
\eeq
where $p'(\frac{\la}{n}) = \partial_{\la} p(\frac{\la}{n})$.
The shift functions can be found from the BAE.  Consider
the equations for vacancies,
\beq
   L\theta(\frac{\lan}{n}) =
      2 \pi I^{n} + \sum_{m=1}^{\infty} \sum_{\g=1}^{N_{m}}
      \Theta_{nm}(\lan - \lgm) + \sum_{\be = 1}^{N_{\La}}
      \theta(\frac{\lan -\La_{\be}}{n})
\eeq
and
\beq
       \sum_{n=1}^{\infty} \sum_{\alpha=1}^{N_{n}}
    \theta(\frac{\La_{\be} - \lan}{n})= 2 \pi J,
\eeq
where $I^{n}$ and $J$ now take on all integers values within the
allowed range.  The BAE after adding one particle to the
$n'$ sea are
\beq
         L\theta(\frac{\tilde{\lan}}{n}) =
      2 \pi (I^{n}+s^{n}) +  \sum_{m=1}^{\infty} \sum_{\g=1}^{N_{m}}
          \Theta_{nm}(\tilde{\la}_{\alpha}^{n} - \tilde{\la}_{\g}^{m})
      + \sum_{\be = 1}^{N_{\La}}
   \theta(\frac{\tilde{\lan} -\tilde{\La}_{\be}}{n})
      + \Theta_{nn'}(\tilde{\lan}-\la_{p})
      -\Theta_{nn'} (\tilde{\lan} -\la_{h})
\eeq
and
\beq
     \sum_{\alpha=1}^{N_{n}}
 \theta(\frac{\tilde{\La}_{\be} - \tilde{\la}_{\alpha}^{n}}{n})
+\theta(\frac{\tilde{\La}_{\be}-\la_{p}}{n'})
-\theta(\frac{\tilde{\La}_{\be}-\la_{h}}{n'})
= 2 \pi (J+s^{\La}).
\eeq
Here $s^{n} \ (n \neq n')$, $s^{\La}$ are arbitrary half integers, $s^{n'}$ is
an integer arising from from a freedom to shift the
excited state distribution by a constant.  Shifting the seas by
by half-integers (integer) leads to an ${\cal O} (1/L)$ shift in
the energy and an ${\cal O} (1)$ shift in the momentum.
Subtracting the shifted equations from the unshifted one and writing
the differences $f(x+dx) -f(x)$ as $f'(x) dx$, one obtains
\begin{eqnarray}
      L f_{n}(\lan)(\lan - \tilde{\la}^{n}_{\alpha})
        & = &
  \sum_{m=1}^{\infty} \sum_{\g=1}^{N_{m}}
A_{nm}(\lan- \la_{\g}^{m})((\lan - \tilde{\la}_{\alpha}^{n})
    -(\la_{\g}^{m} - \tilde{\la}_{\g}^{m})) \\
        & + &
    \sum_{\beta=1}^{N_{\La}} f_{n} (\lan -\La_{\beta})
 ((\lan - \tilde{\la}_{\alpha}^{n}) - (\La_{\beta} - \tilde{\La}_{\beta})) \\
    & - & {1\over2\pi}\Theta_{nn'}(\tilde{\la}_{\alpha}^{n} -\la_{p})
      - s^{n}
\end{eqnarray}
and
\beq
      \sum_{n=1}^{\infty} \sum_{\alpha=1}^{N_{n}}
f_{n}(\La_{\beta} - \lan)
  ((\la_{\be} - \tilde{\la}_{\be})-(\lan -\tilde{\la}_{\alpha}^{n}))
- {1\over2\pi} \th(\frac{\tilde{\La}_{\beta} -\la_{p}}{n'})
   = -s^{\La}
\eeq
Then using ~(\ref{tpbc1}), ~(\ref{tpbc2}) one obtains the following
integral equations for the shift functions:
\begin{eqnarray}
\label{sn}
           S_{n}(\la) & = & -\sum_{m=1}^{\infty}
    \int d\nu A_{nm}(\la-\nu) \theta_{m}(\nu)S_{m}(\nu) \\ \nonumber
   &-&\int d\La f_{n}(\la - \La) \theta_{\La}(\La) S_{\La}(\La) \\
 \nonumber   & -& {1\over2\pi} \Theta_{nn'}(\la-\la_{p})
        -  s^{n}
\end{eqnarray}
and
\beq
\label{sl}
S_{\La}(\La) = \sum_{n=1}^{\infty} \int d\nu
f_{n}(\La-\la) \theta_{n}(\la) S_{n}(\la)
+ {1\over2\pi} \theta(\frac{\La -\la_{p}}{n'})+ s^{\La}
\eeq
We can now show an interesting identity.  First rewrite
\begin{eqnarray}
     - {1 \over 2 \pi} \Theta_{nn'}(\la-\la_{p}) &=&
    -\int^{\la_{p}}_{-\infty} d\nu
     {1\over 2\pi}(\partial_{\nu} \Theta_{nn'})(\la-\nu)
        - {1 \over 2 \pi} \Theta_{nn'}(\infty) \\
      & =&    \int_{-\infty}^{\la_{p}} d\nu A_{nn'}(\la-\nu)
             - \half (2 {\rm min} (n,n') - \delta_{nn'} -1),
\end{eqnarray}
and also
\beq
{1\over2\pi} \theta(\frac{\La -\la_{p}}{n'}) =
 -\int_{-\infty}^{\la_{p}} d\nu e_{n'} (\la-\nu) +\half .
\eeq
Taking the derivative of the integral equations for the functions
$\eps_{n}$,$\eps_{\La}$ ((\ref{te1}) and (\ref{te2})), with respect to
$\la$,$\La$ and noting that all of the kernels are symmetric
functions, one can see by simple substitution that
\beq
      \Delta E = \eps_{n'}(\la_{p}) -B(2n'-1) + \mu.
\eeq
In other words, the functions $\eps_{n}$ may be interpreted
as the dressed excitation energy above the equilibrium.  Proceeding
in a similar way for particle-hole excitation in the $\La$ sea,
one obtains
\beq
       \Delta E = \eps_{\La}(\La_{p}) -B -\mu
\eeq
and
\beq
\label{pl}
        \Delta P = - \sum_{n=1}^{\infty} \int d\la
     p'(\frac{\la}{n}) \th_{n}(\la) \bar{S}_{n}(\la)
      \equiv k_{\La} (\La_{p}),
\eeq
where the shift functions $\bar{S}_{n}(\la|\La_{p})$ and
$\bar{S}_{\La}(\La|\La_{p})$ satisfy
\begin{eqnarray}
\label{bsn}
          \bar{S}_{n}(\la) &=& -\sum_{m=1}^{\infty} \int d\nu
    A_{nm}(\la-\nu) \theta_{m}(\nu)\bar{S}_{m}(\nu) \\ \nonumber
    &-& \int d\La f_{n}(\la-\La) \th_{\La}(\La) \bar{S}_{\La}(\La)
 \\  \nonumber &-&  {1\over2\pi} \theta(\frac{\la -\la_{p}}{n})-
        \bar{s}^{n}
\end{eqnarray}
and
\beq
\label{bsl}
   \bar{S}_{\La}(\La) =  \sum_{n=1}^{\infty} \int d\la
    f_{n}(\La-\la) \theta_{n}(\la) \bar{S}_{n}(\la)
    +\bar{s}^{\La}.
\eeq

Hole excitations may also be constructed.  The equations for the shift
functions are the same as~(\ref{sn}),(\ref{sl}), except the
inhomogeneous terms have opposite sign.. This leads to the energy
\beq
     \Delta E = -\eps_{n'}(\la_{h}) + B(2n'-1) + \mu.
\eeq
Composite excitations are sums of the elementary ones, for instance
a particle-hole excitation in the $n'$ sea has shift functions
\beq
          S_{n}(\la|\la_{p},\la_{h}) =
   S_{n} (\la|\la_{p}) + S_{n} (\la|\la_{h}),
\eeq
A general order one excitation above the equilibrium is of the form
\beq
     \Delta E = \sum_{n=1}^{\infty} \sum_{k} \eps_{n}(\la_{p_{k}})
                -\sum_{n=1}^{\infty} \sum_{l} \eps_{n} (\la_{h_{l}})
                + \sum_{m} \eps_{\La}(\La_{p_{m}})
                - \sum_{n} \eps_{\La}(\La_{h_{n}}),
\eeq
with momentum
\beq
         \Delta P = \sum_{n=1}^{\infty} \sum_{k} k_{n}(\la_{p_{k}})
                -\sum_{n=1}^{\infty} \sum_{l} k_{n} (\la_{h_{l}})
                + \sum_{m} k_{\La}(\La_{p_{m}})
                - \sum_{n} k_{\La}(\La_{h_{n}}).
\eeq

An interpretation of these results is as follows~\cite{yang}: Consider
the functions $\eps_{n}(\la,\mu,T)$ and $\eps_{\La}(\la,\mu,T)$ as
temperature dependent excitation energies of a {\it noninteracting}
fermi system.  If one calculates the free energy of such a system, the
result is exactly the answer for the true free energy ~(\ref{free}) of
the interacting system. There is a freedom to choose how many
species of fermions represent the system, either an infinite number for
equation ~(\ref{free}) or two for equation ~(\ref{free}).

\section{Ground State and Excitations}
\setcounter{equation}{0}
\subsection{Ground State}
We shall obtain the ground state and excitations by taking the $T
\rightarrow 0$ limit of the thermodynamic equations. Taking the zero
temperature limit of equations ~(\ref{te1}) and ~(\ref{te2}), we note
that $\eps_{n} \geq 0$ for $n \geq 2$.  This makes the dependence of
the integral equations on $\eps_{n}$, for  $n \geq 2$, vanish in the zero
temperature limit yielding the dressed energies
\beq
\label{e1}
           \eps_{1}(\la) = \eps^{0}_{1} (\la) + \int d\La f_{1} (\la - \La)
                       \eps_{\La}^{-}(\La),
\eeq
\beq
\label{e2}
           \eps_{\La}(\La)= -(\mu + B) - \int d\la f_{1}(\La-\la)
           \eps_{1}^{-}(\la),
\eeq
and
\beq
\label{e3}
            \eps_{n}(\la) = \eps^{0}_{n}(\la) +
          \int d\La f_{n} (\la -\La)\eps_{\La}^{-}(\La) -
          \int d\nu f_{n-1} (\la-\nu) \eps_{1}^{-} (\nu),
\eeq
where
\beq
     \eps^{-} (\la) = \cases{ \eps(\la) & if $\eps < 0$ \cr
                                   0    & if $\eps \geq 0$. \cr}
\eeq
This tells us that the ground state consists of two fermi seas:
one of $N_{1}$ parameters and one of $N_{\La}$ parameters.
This contrasts with Lai's ansatz which yields a ground state
consisting of spin up electrons and and bound state singlet
electron pairs.


In spectral parameter space, the seas are filled according to
Figure~2.  The $N_{1}$ sea is filled symmetrically around $\la =0$ up
to some fermi level $ \pm Q_{1}$. The $N_{\La}$ sea is is filled
placing particles at the maximum spectral parameter down to a fermi
level $Q_{\La}$.

The ground state energy is
\beq
        \frac{E_{0}}{L} = 1- \int_{-Q_{1}}^{Q_{1}} d\la
           4 \pi f_{1}(\la) \rho_{1}^{p}(\la),
\eeq
where the densities $\rho_{1}^{p}$ and $\s^{p}$ satisfy the
integral equations
\beq
         \rho_{1}^{p}(\la)= f_{1}(\la) -
  (\int_{-\infty}^{-Q_{\La}} + \int_{Q_{\La}}^{\infty}) d\La
      f_{1}(\la-\La) \s^{p} (\La)
\eeq
and
\beq
          \s^{p}(\La)= \int_{-Q_{1}}^{Q_{1}} d\la
       f_{1} (\La - \la) \rho_{1}^{p}(\la).
\eeq
The fermi boundaries $Q_{1}$ and $Q_{\La}$ may be determined from
\beq
        D=\frac{N_{e}}{L} = 1-  \int_{-Q_{1}}^{Q_{1}} d\la \rho^{p}_{1}(\la)
          +  (\int_{-\infty}^{-Q_{\La}} + \int_{Q_{\La}}^{\infty}) d\La
             \s^{p}(\La)
\eeq
and
\beq
          \frac{2 S_{z}}{L} = 1-  \int_{-Q_{1}}^{Q_{1}} d\la \rho^{p}_{1}(\la)
          -  (\int_{-\infty}^{-Q_{\La}} + \int_{Q_{\La}}^{\infty}) d\La
             \s^{p}(\La).
\eeq

In zero magnetic field, the magnetization density of the ground state
is zero, due to a theorem by Lieb and Mattis~\cite{lieb}.  This
requires that $N_{1} = L - N_{\La}-1$.  The number of available spaces
for $N_{1}$ parameters according to~(\ref{max1}) is $N_{1}$, so the
sea is entirely filled.  Thus the fermi sea boundary $Q_{1}$
goes to $\infty$ in the thermodynamic limit.

Considering the half-filled case, $N/L = 1$, in zero magnetic field, we
have $Q_{\La} = 0$ and $Q_{1} = \infty$.  Here the equations
for the ground state are easily solved by fourier transform.
The result for the ground state energy is $E_{0} = 1- 2 \log 2 $,
in agreement with~\cite{schlottman} and~\cite{bares}.

\subsection{Excitation Spectrum in Zero Magnetic Field}

We shall consider order one excitations spectrum in zero magnetic
field. The dressed energy of the excitations is given by the functions
$\eps_{1}$, $\eps_{\La}$ satisfying (\ref{e1}) and (\ref{e2}).
Consider the case of $L$ even, $N_{1}$ odd, and $N_{\La}$ even.  The
momenta of an excitation in the $N_{1}$ sea is given by the zero
temperature limit of~(\ref{pn}),(\ref{sn}),and (\ref{sl}):
\beq
\label{ps}
            k_{1} (\la_{p}) = p(\la_{p}) -\int_{-\infty}^{\infty} d\nu
            p'(\nu) S_{1}(\nu|\la_{p})
\eeq
where $S_{1}$,$S_{\La}$ satisfy the equations
\beq
        S_{1}(\la|\la_{p}) = -(\int_{-\infty}^{-Q_{\La}} +
    \int_{Q_{\La}}^{\infty}) d\La f_{1} (\la-\La) S_{\La}(\La),
\eeq
and
\beq
        S_{\La}(\La|\la_{p}) = \int_{-\infty}^{\infty} d\la
        f_{1}(\La-\la) S_{1} (\la) - {1 \over 2 \pi}\th(\La-\la_{p}).
\eeq
Excitations in the $N_{\La}$ sea have momentum given by
the zero temperature limit of~(\ref{pl}),(\ref{bsn}), and (\ref{bsl}):
\beq
 k_{\La}(\La_{p})  =  - \int_{-Q_{1}}^{Q_{1}} d\la
            p'(\la) \bar{S}_{1}(\la|\La_{p}),
\eeq
where the shift functions $ \bar{S}_{1}$, $\bar{S}_{\La}$ satisfy
\beq
        \bar{S}_{1} (\la|\La_{p}) =-(\int_{-\infty}^{-Q_{\La}} +
    \int_{Q_{\La}}^{\infty} d\La f_{1})(\la-\La) \bar{S}_{\La}(\La|\La_{p})
       + {1\over 2\pi} \th(\la-\La_{p}),
\eeq
\beq
\label{pe}
      \bar{S}_{\La} (\La|\La_{p}) = \int_{-\infty}^{\infty} d\la
            f_{1}(\La -\la) \bar{S}_{1} (\la|\La_{p}).
\eeq
For the case under considiration, all the shifts $s^{1}$,$s^{\La}$,
$\bar{s}^{1}$, and $s^{\La}$ are zero.
The momentum of adding arbitrary numbers of particles and
holes to the fermi seas is
\beq
       \Delta P = \sum_{N_{1} \ particles} k_{1} (\la_{p})
                  - \sum_{N_{1} \ holes} k_{1}(\la_{h})
                 + \sum_{N_{\La} \ particles} k_{\La}(\La_{p})
                  - \sum_{N_{\La} \ holes} k_{\La}(\La_{h}).
\eeq

To see how these functions combine into excitations, first consider
the less than half filled case, $D <1$. The possible elementary
processes are:

\begin{enumerate}

\item Create hole in $N_{1}$ sea:
  $ \Delta N_{1} = -1$, $\Delta N_{\La} = 0$.
This corresponds to adding a spin up electron,
$\Delta N_{e} = 1 \ \ \Delta S= \half$ .  The
energy and momentum are
\beq
      \Delta E = -\eps_{1}(\la_{h}), \qquad  \Delta P = -k_{1} (\la_{h}).
\eeq

\item Create hole in $N_{\La}$ sea:
$ \Delta N_{1} = 0$, $ \Delta N_{\La} = -1$.
This corresponds to removing a spin down electron,
 $\Delta N_{e} = -1$, $\Delta S = \half$.
This creates a hole in the $N_{1}$ sea, so the energy and
momentum are
\beq
          \Delta E = -\eps_{1} (\la_{h}) -\eps_{\La} (\La_{h}), \qquad
           \Delta P = -k_{1}(\la_{h}) - k_{\La}(\La_{h}).
\eeq

\item Transfer particle from  $N_{\La}$ sea to $N_{1}$ sea:
$\Delta N_{1} = +1$, $ \Delta N_{\La} = -1$. Removing a $\La$
parameter increases holes in the $N_{1}$ sea by one, in which the
one $\la$ parameter is placed.  Thus, there is no parametric
dependence on $\la$. This excitation removes 2 electrons, with no
change in spin, $\Delta N_{e} = -2$, $\Delta S = 0$. The energy and
momenta are
\beq
           \Delta E = -\eps_{\La}(\La_{h}), \qquad
          \Delta P = -k_{\La}(\La_{h}).
\eeq

\item Transfer particle from $N_{1}$ sea to $N_{\La}$ sea:
$\Delta N_{1} = -1$, $\Delta N_{\La} = +1$.
As in type 3, there is no parametric $\la$ dependence.
This excitation adds 2 electrons, with no change in spin.
The energy and momentum of the excitation are
\beq
     \Delta E = \eps_{\La}(\La_{p}), \qquad
      \Delta P = k_{\La} (\La_{p}).
\eeq

\item Bind particles into bound state:
$\Delta N_{1} = -2$, $\Delta N_{\La} = -1$, $ \Delta N_{2} = + 1$.

This opens up two holes in the $h+\down$ sea, and one in the $\down$
sea.  There is only one vacancy for the bound state, so there is no
parametric dependence on its location.  This excitation does not change
the number of electrons or the spin. The energy and momentum are
\beq
  \Delta E =
 -\eps_{1}(\la_{h_{1}}) - \eps_{1}(\la_{h_{2}}) -\eps_{\La}(\La_{h}),
 \qquad \Delta P =
   -k_{1}(\la_{h_{1}}) - k_{1}(\la_{h_{2}})-k_{\La}(\La_{h}).
\eeq

\end{enumerate}

\subsection{Excitations at Half Filling}

In the half filled case, $D=1$,  we can solve the integral equations
 (\ref{e1}),(\ref{e2}),(\ref{ps})-(\ref{pe}) for
the energy and momenta.  The dressed energy and momentum of
an excitation in the $N_{1}$ sea are
\beq
         \eps_{1} (\la) = \frac{-\pi}{\cosh(\frac{\pi \la}{2})}
\eeq
and
\beq
          k_{1} (\la) = -\tan^{-1} (\sinh(\frac{\pi \la}{2})) + {\pi \over 2}.
\eeq
For an excitation in the $N_{\La}$ sea, the dressed energy and momentum are
\beq
       \eps_{\La} (\La) = R(\La) - 2 \log 2
\eeq
and
\beq
              k_{\La}(\La) = \int_{-\infty}^{-\La} d\La' R(\La'),
\eeq
where
\beq
           R(\La) = \int_{-\infty}^{\infty} \frac{e^{-i \La w}}
            {1+e^{2 |w|}}.
\eeq
The $n$-string excitations have zero energy, $\eps_{n>1} =0$.
The ground state has both fermi seas filled,
$N_{1} = L/2$, $N_{\La} = L/2 -1$.

These functions combine to form excitations in the following ways:

\begin{enumerate}

\item Create hole in $N_{\La}$ sea:
$ \Delta N_{1} = 0$, $ \Delta N_{\La} = -1$.
This corresponds to removing a spin down electron,
 $\Delta N_{e} = -1$, $\Delta S = \half$.
This creates a hole in the $h+\down$ sea, so the energy and
momentum are
\beq
          \Delta E = -\eps_{1} (\la_{h}) -\eps_{\La} (\La_{h}) \qquad
           \Delta P = -k_{1}(\la_{h}) - k_{\La}(\La_{h}).
\eeq

\item Transfer particle from  $N_{\La}$ sea to $N_{1}$ sea:
$\Delta N_{1} = +1$, $ \Delta N_{\La} = -1$ In contrast with
the case away from half filling, now two holes in
the $N_{\La}$ sea are created. This excitation
removes 2 electrons, with no change in spin,
$\Delta N_{e} = -2$, $\Delta S = 0$. The energy and momenta are
\beq
   \Delta E = -\eps_{\La}(\La_{h_{1}}) - \eps_{\La}(\La_{h_{2}}), \qquad
          \Delta P = -k_{\La}(\La_{h_{1}}) -k_{\La}(\La_{h_{2}}).
\eeq

\item Create holes in $N_{1}$ and $N_{\La}$ seas:
 $\Delta N_{1} = -1$, $\Delta N_{\La} = -1$. This is not simply
a sum of type 1 and 2 excitations for $D<1$ because here there
is no hole in the $\down$ sea. This excitation has
$\Delta N_{e} = 0$,$\Delta S = 1$.  The energy and momentum are
\beq
             \Delta E = -\eps_{1}(\la_{h_{1}}) - \eps_{1}(\la_{h_{2}}),\qquad
               \Delta P = -k_{1}(\la_{h_{1}}) - k_{1}(\la_{h_{2}}).
\eeq

\item Bind particles into bound state:
$\Delta N_{1} = -2$, $\Delta N_{\La} = -1$, $ \Delta N_{2} = + 1$.
This opens up two holes in the $N_{1}$ sea, and none in the $N_{\La}$
sea, in contrast with type 5 excitation for $D<1$.  There is only one
space for the bound state, so there is no parametric dependence its
location.  This excitation has $\Delta N_{e} = 0$, $\Delta S = 0$.
The energy and momentum are
\beq
  \Delta E = -\eps_{1}(\la_{h_{1}}) - \eps_{1}(\la_{h_{2}}), \qquad
               \Delta P = -k_{1}(\la_{h_{1}}) - k_{1}(\la_{h_{2}}).
\eeq

\end{enumerate}

The excitations 3 and 4 are identical to those in the $XXX$ spin
chain~\cite{faddeev}.  The results here agree with those obtained
in~\cite{bares} using Lai's and Sutherland's ansatz.

\section*{Acknowledgements}

The author had useful discussions with F.H. Essler, R. Kedem, V.E.
Korepin, and B.M. McCoy.  The author thanks V.E. Korepin for
suggesting the problem.

\begin{appendix}
\setcounter{equation}{0}
\section{Relationship to Lai's Ansatz}

In this section we shall how the solutions of Essler and Korepin's
ansatz, (\ref{pbc1}) and (\ref{pbc2}), are related to solutions of
Lai's ansatz (\ref{l1}) and (\ref{l2}).  A mathematical equivalence
between the ansatzes has been shown in~\cite{essler} and
{}~\cite{bares2}.

First we recall the Lai solution~\cite{schlottman,lai} which
treats the electrons as dynamical objects in a background of
empty sites.  The $N_{\down}+N_{\up}$ spectral parameters $\la_{l}$
describe the kinetic degrees of freedom, the $N_{\down}$ parameters
describe the motion of the spins of the electrons.  The
spectal parameters must satisfy the BAE:
\beq
\label{l1}
        {\bl \frac{\la_{l} +i}{\la_{l} -i} \br}^{L} =
  \prod_{\beta =1}^{N_{\down}}
   \bl \frac{\la_{l} -\La_{\beta} + i}{\la_{l} -\La_{\beta} - i} \br
\ \ \ \ \ \ l=1,...,N_{\down}+N_{\up},
\eeq
\beq
\label{l2}
      \prod_{l=1}^{N_{\down}+N_{\up}}
   \bl \frac{\La_{\beta} - \la_{l} + i}{\La_{\beta} - \la_{l} -i} \br=
   - \prod_{\gamma=1}^{N_{\down}}
    \bl \frac{\La_{\beta} - \La_{\g} + 2 i}{\La_{\beta} - \La_{\g} - 2i} \br.\
     \ \ \ \ \beta=1,...,N_{\down}.
\eeq

In this ansatz, there are two types of string solutions:
\begin{enumerate}
\item  bound singlet electron pairs : composed of 2 $\la$ and 1 $\La$
parameters:
\beq
       \la = \La +i \ \ \ \ \ \la' = \La -i
\eeq
\item  spin bound states: take $n \ \La$ parameters and set
\beq
             \La_{\delta} = \La + i (n+1-2\delta) \ \ \ \ \
             \delta = 1,...,n.
\eeq
\end{enumerate}
In the thermodynamic limit one define densities $\rho$,$\s'$,
$\s_{1}$, and $\s_{n}$ ($n=2,...,\infty$) for free electrons, bound
pairs, down spins, and spin bound states respectively.  The thermodynamic
limit of the Lai BAE is
\beq
\label{tl1}
   \rho^{t}(\la) = f_{1}(\la) - \int d\La
 f_{1}(\la-\La) \s^{'p}(\La) - \sum_{n=1}^{\infty}
     \int d\nu f_{n}(\la-\nu) \s_{n}^{p}(\nu),
\eeq
\beq
\label{tl2}
     \s^{'t}(\La)  = f_{2} (\La) - \int d\La'
  f_{2} (\La-\La') \s^{'p}(\La') -
  \int d\la f_{1}(\La-\la) \rho^{p}(\la),
\eeq
and
\beq
\label{tl3}
      \s_{n}^{t}(\la) = \int d\nu f_{n}(\la-\nu) \rho^{p}(\nu)
     - \sum_{m=1}^{\infty} \int d\nu B_{nm} (\la-\nu)
      \s_{m}^{p}(\nu),
\eeq
where
\beq
       B_{nm} =
\cases{ f_{|n-m|} + 2 f_{|n-m|+2} + ... + 2 f_{n+m-2} + f_{n+m}
 & if $n \neq m$ \cr
        2 f_{2} + ... + 2 f_{2n-2} + f_{2n} & if $n=m$ \cr}.
\eeq

Now if one makes identification between the densities in the
two ansatzes
\beq
     \rho^{p,h} \leftrightarrow \rho_{1}^{h,p} \ \ \ \
     \s'^{p,h}  \leftrightarrow \s^{p,h}  \ \ \ \
     \s_{n}^{p,h} \leftrightarrow \rho_{n+1}^{p,h}, \ n=1,...,\infty,
\eeq
we may see that equations ~(\ref{tl1}), ~(\ref{tl2}), and ~(\ref{tl3})
{it coincide} with ~(\ref{tpbc1}) and ~(\ref{tpbc2}). To show this one
only needs the identity
\beq
          \hat{f}_{n} \hat{f}_{m} = \hat{f}_{n+m},
\eeq
where have written the action of the kernel
as an integral operator on some function $g(\la)$:
\beq
         \hat{f}_{n} g(\la) \equiv \int_{-\infty}^{\infty}
          f_{n} (\la-\nu) g(\nu).
\eeq

Thus the role of particles and hole solutions are interchanged in
the $\up + \down$ sea in Lai's ansatz and the $N_{1}$ sea
in Essler and Korepin's ansatz.  One expects that this will hold
in a finite box as well, the hole solutions of ~(\ref{pbc2})
will coincide with the particle solutions for bound pairs in
Lai's ansatz.

Using this particle-hole correspondence, one can show that the free
energy obtained here is equal to that in ref.~\cite{schlottman}.
Identifying the energy functions, one can rewrite the integral
equations for the $\eps$ functions, (\ref{te1}) and (\ref{te2}), to
coincide with the corresponding equations in Lai's ansatz,equation
(5.16) in~\cite{schlottman}.
\end{appendix}

\end{document}